\documentclass[pdflatex,sn-nature]{sn-jnl}

\usepackage{graphicx}%
\usepackage{multirow}%
\usepackage{amsmath,amssymb,amsfonts}%
\usepackage{amsthm}%
\usepackage{mathrsfs}%
\usepackage[title]{appendix}%
\usepackage{xcolor}%
\usepackage{textcomp}%
\usepackage{manyfoot}%
\usepackage{booktabs}%
\usepackage{algorithm}%
\usepackage{algorithmicx}%
\usepackage{algpseudocode}%
\usepackage{listings}%
\usepackage{bm}%
\usepackage{mathrsfs}%

\theoremstyle{thmstyleone}%
\theoremstyle{thmstyletwo}%

\theoremstyle{thmstylethree}%

\begin{document}
\title[Phonon-assisted free-electron recoil]
{Phonon-assisted free-electron recoil control and mechanical sensing}

\author[1,2]{\fnm{Sunjun} \sur{Hwang}}

\author[1]{\fnm{Lee Wei Wesley} \sur{Wong}}

\author*[1]{\fnm{Liang Jie} \sur{Wong}}\email{liangjie.wong@ntu.edu.sg}

\author*[1,2,3]{\fnm{Leong Chuan} \sur{Kwek}}\email{kwekleongchuan@nus.edu.sg}

\affil*[1]{\orgdiv{School of Electrical and Electronic Engineering}, \orgname{Nanyang Technological University}, \orgaddress{\street{50 Nanyang Avenue}, \city{Singapore}, \postcode{639798}, \country{Singapore}}}

\affil[2]{\orgdiv{Centre for Quantum Technologies}, \orgname{National University of Singapore}, \orgaddress{\street{3 Science Drive 2}, \city{Singapore}, \postcode{117543}, \country{Singapore}}}

\affil[3]{\orgdiv{Natural Sciences and Science Education}, \orgname{National Institute of Education}, \orgaddress{\street{1 Nanyang Walk}, \city{Singapore}, \postcode{637616}, \country{Singapore}}}

\abstract{Mechanical vibrations are fundamental to a wide range of physical phenomena and technologies, yet their ability to actively shape free-electron dynamics remains largely unexplored. Here we theoretically show how an optomechanical membrane provides a phonon-assisted optical interface for controlling free-electron recoil and sensing localized forces. Transverse membrane motion scatters cavity light into high-momentum evanescent sidebands that enable phase-matched interaction with electrons traveling in vacuum. Beyond supplying momentum for phase matching, the vibration amplitude controls the electron–light interaction strength and redistributes electron populations among recoil orders. External forces that perturb the vibration are thereby encoded in the electron recoil spectrum. Under the modeled conditions, electron readout of localized stochastic forces can yield a higher signal-to-noise ratio than a Gaussian optical homodyne benchmark at equal detected probe rates. These results provide a theoretical basis for mechanically reconfigurable free-electron optics and mode-selective mechanical sensing.
}


\maketitle

\section*{Overview and motivation}

Mechanical vibrations are ubiquitous, from macroscopic structures to
nanoscale devices. Their interaction with light underpins cavity
optomechanics, where radiation pressure connects mechanical motion to
an optical field \cite{Aspelmeyer2014,bibOpt0}. This connection has
enabled precision measurements approaching quantum limits
\cite{bibOpt6,bibOpt2,Huang2024}, as well as the transfer of
information between mechanical and electromagnetic systems
\cite{bibOpt8,bibOpt9,Jiang2020}. By allowing light both to drive
motion and to detect its response, optomechanics makes mechanical
resonators useful intermediaries between distinct physical systems.
Extending this role to freely propagating charged particles may offer
a route for mechanical motion both to control a particle state and to
transfer information to it.

Free electrons provide a natural setting for this extension. Beyond
their traditional role in microscopy, they can be coherently
manipulated by light on ultrafast and nanometer scales, including
through quantized energy exchange in photon-induced near-field
interactions \cite{bibPINEM1,bibPINEM2}. Subsequent work demonstrated
coherent optical phase manipulation \cite{Feist2015,Ropers2016},
enhanced coupling to optical resonators
\cite{bibPINEM11,Wang2020,Henke2021,bibPINEM23}, stimulated
interactions in nanoscale plasmonic fields \cite{bibPINEM19}, and
control of transverse and longitudinal electron degrees of freedom
\cite{bibKD6,bibKD8,bibPINEM24,bibKD2}. These advances have
transformed the free electron from a passive probe into a controllable
matter wave and, increasingly, a quantum-optical resource for
manipulating and sensing light--matter systems
\cite{bibPINEM9,bibFree1,bibPINEM12,bibPINEM18}.

A recurring requirement in these free-electron--light interactions is
phase matching. Optical contributions encountered along the electron
trajectory must accumulate coherently rather than cancel. Evanescent near fields generated at micro- and nanophotonic structures can provide the momentum required for phase-matched coupling~\cite{Dahan2020,Kozak2017,Talebi2020}. Strong optical fields in free space provide complementary routes to electron manipulation through ponderomotive interactions~\cite{bibKD7,bibKD5}.
These approaches have established powerful routes to electron--light
coupling, but the interaction is generally determined by the electron
parameters together with a prescribed optical environment.
Introducing a mechanical degree of freedom changes this picture: the
electromagnetic environment can itself respond to material motion,
linking the electron--light interaction to a dynamical mechanical
coordinate.

This connection suggests a possible role for free electrons in mechanical sensing. Conventional optical displacement measurements
sample mechanical motion over a finite optical area, so their response
depends on how the optical field overlaps with the mechanical mode
shape \cite{Haghighi2018}. For spatially structured or higher-order modes, regions moving in
opposite directions can partially cancel in the detected signal, making an appreciable vibration weakly visible or nearly dark to a particular optical probe \cite{Haghighi2018,Roxworthy2018}. Consequently, the optical accessibility of such modes can depend strongly on spatial mode matching and probe position, motivating localized or spatially resolved readout schemes
\cite{Iyer2024,Roxworthy2018}. At the same time, the spatial structure
of mechanical modes carries information about how and where a force drives the resonator: the spatial profile of an applied force
determines how strongly individual mechanical modes are excited and
therefore provides a basis for mode-selective force sensing \cite{Esmenda2021}. This motivates a complementary probe that samples the motion without averaging over the same regions of opposite displacement. Because free-electron interactions can depend coherently on the phase accumulated along the electron trajectory, free electrons may provide such an alternative sensing approach \cite{Dahan2020,Kozak2017}.

Here we develop a framework in which a vibrating membrane acts as a
dynamically reconfigurable optical boundary for a passing free
electron. The mechanical motion generates the high-spatial-frequency
optical near field required for phase matching, while its amplitude
determines the strength of the recoil interaction. This allows the
mechanical degree of freedom to serve not only as a means of satisfying
momentum matching but also as a control variable for the outgoing
electron state. We then use the same interaction to read out external
perturbations: a localized force modifies the mechanical motion, and
the resulting change is transferred to the electron recoil spectrum.
We analyze optical preparation of the required high-order mechanical
modes, the mode selectivity obtained by tuning the electron energy,
and the resulting force sensitivity. Comparison with Gaussian optical
homodyne readout shows that electron detection can remain sensitive to
mechanical modes whose optical signal is strongly suppressed by
spatial averaging. Together, these results establish mechanical motion as an active input to free-electron optics by showing how controlled mechanical excitation shapes the electron recoil distribution.

\subsection*{Mechanical control of free-electron recoil}

\begin{figure}[!t]
\centering
\centerline{\includegraphics[width=\textwidth]{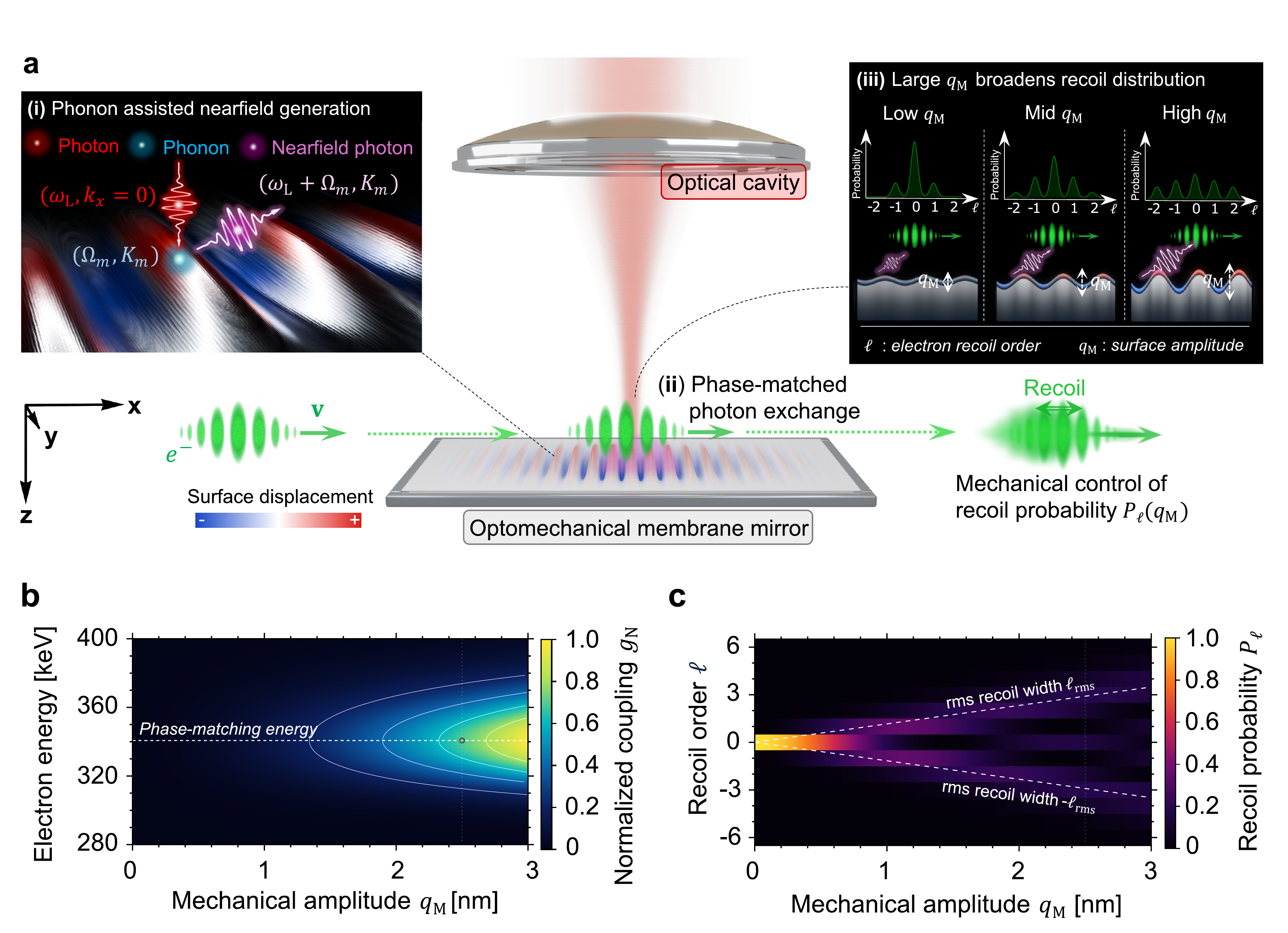}} \caption{\textbf{Mechanical control of free-electron coupling and recoil.}
\textbf{a}, A free electron travels through vacuum above an optically
illuminated membrane supporting a transverse mechanical mode.
\textbf{(i)}, An optical carrier at frequency $\omega_{\rm L}$ scatters
through interaction with a membrane phonon, generating an evanescent
optical sideband at $\omega_{\rm L}+\Omega_m$ with tangential wave
number $K_m$.
\textbf{(ii)}, The mechanically generated near field carries the large
tangential wave number $K_m$, supplying the momentum required to satisfy
the electron--light phase-matching condition.
\textbf{(iii)}, The amplitude of this generated near field scales with
the membrane displacement $q_{\rm M}$, allowing the strength of the
phase-matched electron interaction to be mechanically controlled and
thereby redistributing the electron population among recoil orders. Schematics are
not to scale.
\textbf{b}, Normalized squared electron coupling
$g_{\rm N}=g_m^2/\max g_m^2$ versus mechanical amplitude and electron
kinetic energy. The coupling is concentrated around the phase-matched
energy and increases with $q_{\rm M}$ without shifting the matching
condition, showing that the mechanical amplitude controls the
interaction strength independently of the matched electron energy.
White contours indicate constant $g_{\rm N}$; the horizontal dashed and vertical dotted lines mark the phase-matched energy $\simeq340.7\,{\rm keV}$ and the reference amplitude $q_{\rm ref}=2.5\,{\rm nm}$, respectively. The point denotes
the reference operating coupling $g_{50}=4.11$.
\textbf{c}, Recoil-order populations at the phase-matched energy as
$q_{\rm M}$ is varied. Increasing the mechanical amplitude redistributes
probability among progressively higher recoil orders while their spacing,
$\hbar K_{m}$, remains fixed. Dashed curves indicate the RMS recoil
width and serve as guides to the characteristic momentum spread.}
\label{fig1}
\end{figure}

We consider an electron traveling through vacuum above an illuminated
square membrane of side length $L$ (Fig.~\ref{fig1}\textbf{a}).
In a tension-dominated fixed-boundary model (see Methods and Supplementary Sec.~S1.2), the $(m,1)$ mechanical mode
has profile $\psi_m(x,y)=\sin(m\pi x/L)\sin(\pi y/L)$ and displacement
$Q_m(x,y,t)=q_{\rm M}\psi_m(x,y)\cos(\Omega_m t)$ normal to the
equilibrium membrane plane at $z=L_{\rm c}$. Here $q_{\rm M}$ is the peak out-of-plane displacement and $\Omega_m$
the mechanical angular frequency. The electron travels nearly parallel to the membrane along $y=L/2$, where
$\psi_m(x)=\sin(m\pi x/L)$.

Illumination at angular frequency $\omega_{\rm L}$ turns the vibrating
surface into a dynamic optical grating \cite{Sipe2016}. For the
approximately normally incident cavity carrier, the deformation
supplies tangential wavevectors $\pm K_m$, with $K_m=m\pi/L$, and
frequency shifts $\pm\Omega_m$. We consider coupling to the upper optical sideband at
$\omega_{\rm L}+\Omega_m$, assuming that the lower sideband is negligible.
The effect of retaining both sidebands is examined in the
Supplementary Information. For the high-order modes $m=49,50$ considered here,
$K_m>(\omega_{\rm L}+\Omega_m)/c$. Thus, this component lies outside
the vacuum light cone and is evanescent. The boundary therefore
generates the large tangential momentum needed for electron coupling
without requiring the electron to enter the membrane. Its spatial wave number $K_m$ is determined by the mechanical mode, whereas its amplitude scales with the mechanical displacement $q_{\rm M}$. These two properties provide separate controls of the electron interaction.

Phase matching requires the optical phase sampled by the electron to remain nearly stationary along its trajectory, so that contributions from different positions add coherently rather than cancel. For electron speed $v$ and trajectory angle $\theta$ measured
from the membrane plane, the phase mismatch is
\begin{equation}
\Delta_m
=
K_m\cos\theta
-
\frac{\omega_{\rm L}+\Omega_m}{v}.
\label{eq:phase_matching}
\end{equation}
At $\Delta_m=0$, the electron remains in phase with the selected
near-field component. Changing its kinetic energy therefore changes
which mechanical spatial pattern is most efficiently addressed.

Mechanical amplitude then controls the strength of this phase-matched
interaction. For a Gaussian interaction envelope, the coupling magnitude
is approximated by
\begin{equation}
g_m
\simeq
g_{\rm ref}
\frac{q_{\rm M}}{q_{\rm ref}}
e^{-\kappa_m d}
\exp\!\left[
-\frac{v^2\Delta_m^2\tau_{\rm int}^2}{4}
\right],
\label{eq:main_electron_coupling}
\end{equation}
where $d$ is the electron--surface separation, $\kappa_m$ is the
evanescent decay constant, and $\tau_{\rm int}$ is the effective
interaction time. The three factors describe mechanical-amplitude
control, evanescent attenuation, and phase-matching selectivity,
respectively. Figure~\ref{fig1}\textbf{b} shows that increasing
$q_{\rm M}$ strengthens the coupling without shifting the matched
energy. For $m=50$, $q_{\rm ref}=2.5\,\mathrm{nm}$ and
$g_{\rm ref}=5.72$ give $g_{50}=4.11$ at $d=70\,\mathrm{nm}$ and a
phase-matched energy of approximately $341\,\mathrm{keV}$.

At this matched energy, mechanical control of the coupling translates
directly into control of the outgoing electron state. For a
quasi-monoenergetic incident electron, the resolved populations are
$P_\ell=J_\ell^2(g_m)$, where $\ell$ labels the recoil order and
$J_\ell$ is a Bessel function of the first kind. Increasing
$q_{\rm M}$ populates higher recoil orders, so that the rms recoil-order width grows as
$\sqrt{\langle\ell^2\rangle}=g_m/\sqrt{2}$
(Fig.~\ref{fig1}\textbf{c}), while the spacing between adjacent orders
remains $\hbar K_m$. In this single-sideband regime, the mechanical phase remains encoded in the complex scattering amplitudes but does not affect the resolved recoil populations considered here. Thus, linear optical
scattering from the moving boundary produces a nonlinear electron response, with the mechanical amplitude controlling the distribution among momentum states.

\begin{figure}[!t]
\centering
\centerline{\includegraphics[width=\textwidth]{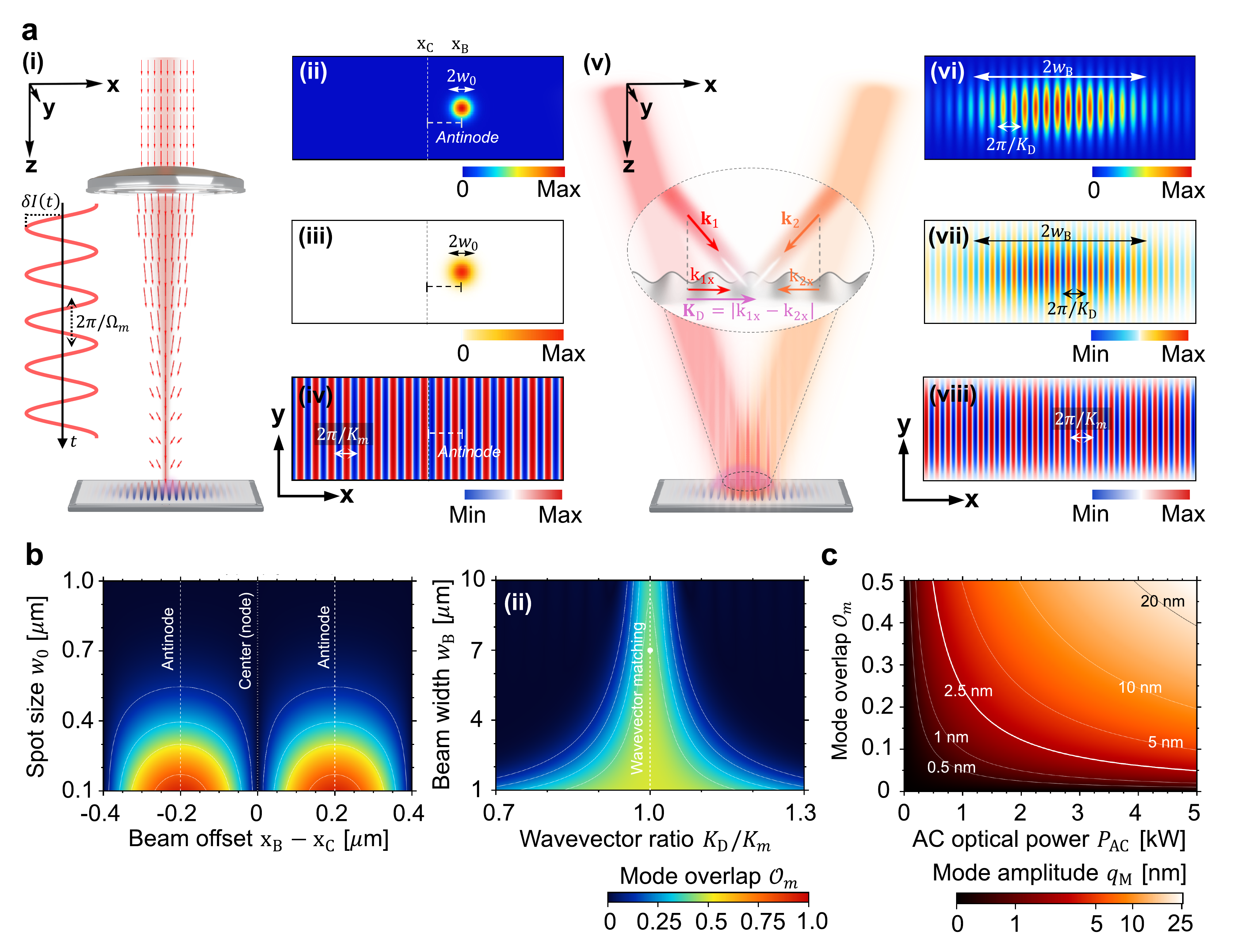}} \caption{\textbf{Optical preparation of high-order transverse mechanical modes.}
\textbf{a}, Two strategies for exciting the standing
$(m,n)=(50,1)$ membrane mode.
\textbf{(i--iv)}, Focused-beam preparation: an intensity-modulated beam
at $\Omega_m$ (\textbf{i}) is localized near a mechanical antinode
(\textbf{ii}), producing an oscillating radiation-pressure profile
(\textbf{iii}) that predominantly drives one displacement region of the
target mode (\textbf{iv}) and thereby reduces spatial cancellation.
\textbf{(v--viii)}, Two-beam preparation: coherent non-collinear beams
with $|\omega_1-\omega_2|=\Omega_m$ (\textbf{v}) generate a traveling
beat-intensity pattern (\textbf{vi}) and corresponding oscillating
radiation pressure (\textbf{vii}). Matching
$K_{\rm D}=|k_{1x}-k_{2x}|$ to $K_m$ aligns this pressure pattern with
the alternating displacement of the target mode (\textbf{viii}),
allowing extended spatial driving without tight focusing.
Intensity and pressure maps are snapshots at a selected drive phase;
schematics are not to scale.
\textbf{b}, Spatial overlap $\mathcal O_m$ for the two preparation
schemes.
\textbf{(i)}, Focused-beam overlap versus spot radius $w_0$ and beam
offset $x_{\rm B}-x_{\rm C}$ from the membrane center $x_{\rm C}=L/2$. The overlap is largest near an antinode
and increases with tighter focusing; dashed and dotted lines mark the
nearest antinodes and central node, respectively.
\textbf{(ii)}, Two-beam overlap versus $K_{\rm D}/K_m$ and
beat-envelope radius $w_{\rm B}$. The overlap peaks near
$K_{\rm D}=K_m$, while broader envelopes narrow the wavevector-matching
range. The white point marks $w_{\rm B}=7\,\mu{\rm m}$.
\textbf{c}, Mechanical amplitude $q_{\rm M}$ versus local optical
modulation power $P_{\rm AC}$ and spatial overlap $\mathcal O_m$.
Larger overlap reduces the optical power required to reach a given
mechanical amplitude, illustrating the benefit of spatially matched
driving. White curves indicate constant $q_{\rm M}$, with
$q_{\rm M}=2.5\,{\rm nm}$ highlighted.}
\label{fig2}
\end{figure}

Exciting this mechanical mode requires both spatial and temporal matching. The standing displacement alternates in sign, so broadly applied
radiation pressure can produce canceling contributions to the selected
mode. Following the force--mode projection principle
\cite{Xi2021,Iyer2024}, we define the normalized overlap
\begin{equation}
\mathcal O_m[\mathcal W]
=
\frac{
\left|\int_{\mathcal A}
\mathcal W(x,y)\psi_m(x,y)\,\mathrm{d}A\right|
}{
\int_{\mathcal A}|\mathcal W(x,y)|\,\mathrm{d}A
}.
\label{eq:main_mode_overlap}
\end{equation}
Here $\mathcal A$ is the membrane area and $\mathcal W$ describes the spatial pattern of the optical drive. For mechanical excitation, $\mathcal W$ represents the oscillating radiation-pressure profile across the membrane. Matching the pressure pattern to the sign and spatial variation of the mechanical displacement reduces cancellation and increases the modal overlap \(\mathcal{O}_m\), so that contributions from different regions add constructively rather than cancel.

A focused beam achieves appreciable overlap by concentrating pressure
near an antinode (Fig.~\ref{fig2}\textbf{a}, i--iv), while intensity
modulation at $\Omega_m$ provides resonant driving. For a Gaussian
intensity profile centered at $(x_{\rm B},L/2)$, the weighting is
$\mathcal W_{\rm F}(x,y)\propto
\exp[-2((x-x_{\rm B})^2+(y-L/2)^2)/w_0^2]$,
where $w_0$ is the $1/e^2$ intensity radius.
For $m=50$, a beam centered on the membrane lies at a mechanical node and therefore gives zero overlap. Displacing the beam by approximately $\pm0.20\,\mu\mathrm{m}$ aligns it with the nearest antinodes and maximizes the overlap (Fig.~\ref{fig2}\textbf{b}(i)). Tighter focusing reduces cancellation
from adjacent regions of opposite displacement. The excitation comes
from localized pressure normal to the membrane, not a net lateral
optical force.

Alternatively, two coherent non-collinear beams pattern the pressure
over an extended region (Fig.~\ref{fig2}\textbf{a}, v--viii).
Their frequency difference $|\omega_1-\omega_2|=\Omega_m$ drives the
resonance, while $K_{\rm D}=|k_{1x}-k_{2x}|$ sets the traveling
pattern's wavevector. A centered Gaussian beat envelope gives
$\mathcal W_{\rm B}(x,y)\propto
\exp[-2((x-L/2)^2+(y-L/2)^2)/w_{\rm B}^2]e^{iK_{\rm D}x}$,
with $w_{\rm B}$ its $1/e^2$ radius. Matching $K_{\rm D}$ to $K_m$
produces an oscillating force on the standing mode across many periods,
without confining the drive to one antinode. The pressure pattern travels, but its projection onto the target standing mode oscillates at the mechanical resonance, thereby driving the standing mode resonantly without requiring a traveling mechanical wave. Figure~\ref{fig2}\textbf{b}(ii) shows a matched overlap of approximately
$0.43$ at $w_{\rm B}=7\,\mu\mathrm{m}$; broader envelopes narrow
the tolerance to wavevector mismatch.

At resonance, displacement scales with overlap and local optical
modulation-power amplitude $P_{\rm AC}$. Under the reflecting-boundary
calibration in Fig.~\ref{fig2}\textbf{c}, increasing the overlap from
$0.10$ to $0.50$ reduces the optical power required at the membrane for
$q_{\rm M}=2.5\,\mathrm{nm}$ from approximately $2.45$ to
$0.49\,\mathrm{kW}$. Spatial matching therefore reduces the
actuation required for a chosen displacement, without requiring the
preparation beam to have the same profile as the subsequent optical
readout.

\subsection*{Mode-resolved electron readout and sensing}

\begin{figure}[!t]
\centering
\centerline{\includegraphics[width=\textwidth]{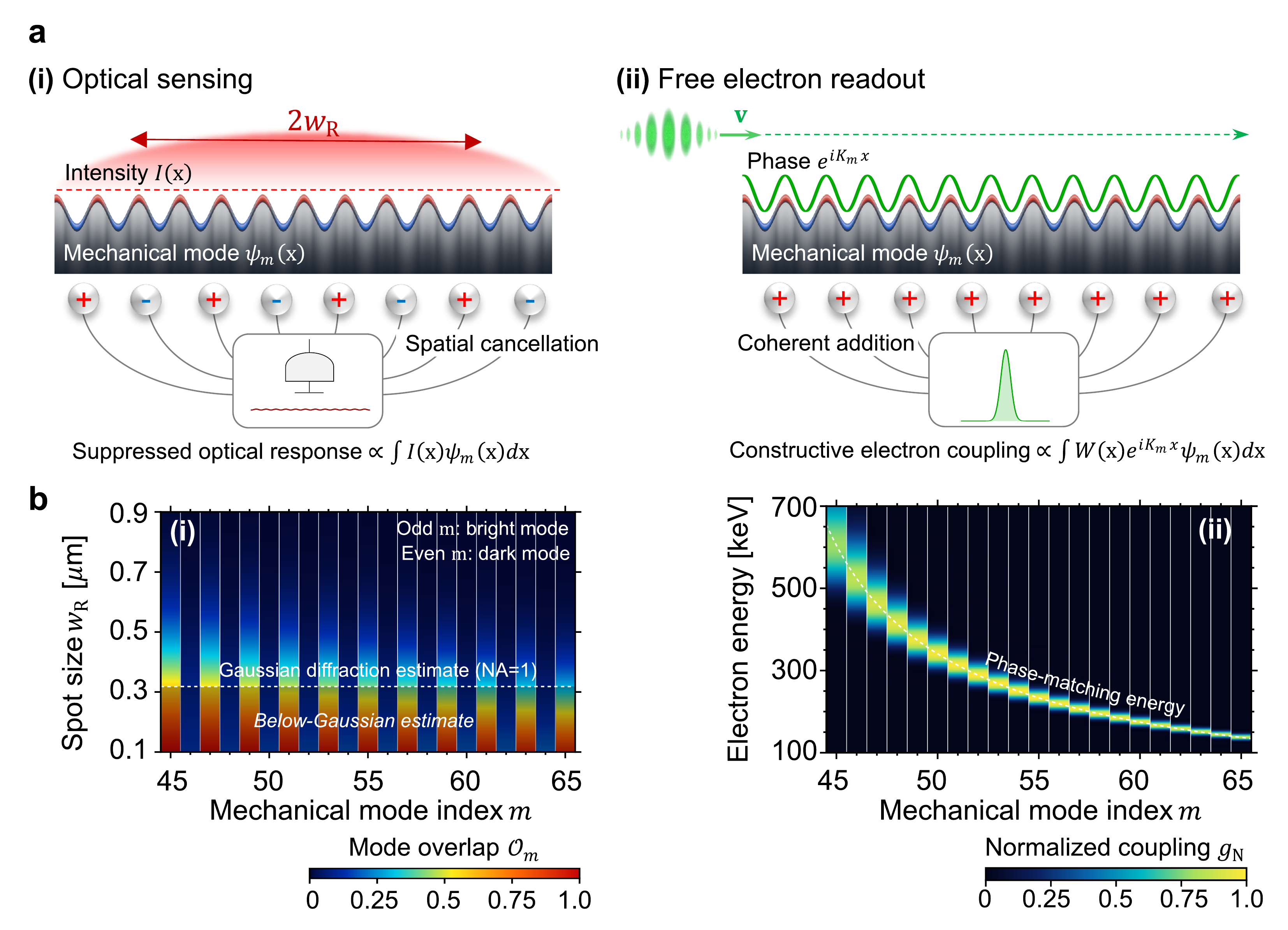}} \caption{\textbf{Momentum-selective free-electron readout of high-order mechanical modes.}
\textbf{a}, Contrasting spatial response of optical and electron readout.
\textbf{(i)}, A Gaussian optical probe averages regions of opposite
mechanical displacement, producing spatial cancellation and a suppressed
optical response.
\textbf{(ii)}, For a phase-matched electron, the interaction phase
compensates the alternating mechanical pattern so that contributions
along the trajectory add coherently. Here $W(x)$ denotes the optical
near-field envelope sampled by the electron; red and blue regions
indicate opposite mechanical displacements. Schematics are not to scale.
\textbf{b}, Calculated readout of mechanical modes $m=45$--$65$.
\textbf{(i)}, Gaussian optical mode overlap $\mathcal O_m$ versus
readout radius $w_{\rm R}$ and mode index. Spatial averaging suppresses
the visibility of high-order modes, with even-$m$ modes particularly
weak near the centered-probe configuration. The dashed line shows the
Gaussian estimate $w_{\rm diff}=\lambda_L/(\pi\,\mathrm{NA})$ for
diffraction-limited focusing at $\mathrm{NA}=1$; the shaded region lies
below this estimate.
\textbf{(ii)}, Normalized squared electron coupling
$g_{\rm N}=|g_m|^2/\max_{m,E}|g_m|^2$ versus mode index and electron
kinetic energy. Phase matching selects different mechanical modes by
electron energy while retaining coupling to both even- and odd-$m$
modes. The dashed curve marks the phase-matching energy, and thin
vertical lines separate individual modes.
}
\label{fig3}
\end{figure}

The overlap defined in Eq.~\eqref{eq:main_mode_overlap} also governs the response of an optical displacement measurement, for which the
detected motion is weighted by the optical intensity \cite{Haghighi2018}. We therefore take $\mathcal W_{\rm R}=I_{\rm G}$ to quantify the optical visibility of each mechanical mode. For a Gaussian probe centered on the membrane, even-$m$ modes are antisymmetric about the center, and their contributions cancel, whereas odd-$m$ modes are symmetric and can
retain a finite optical response. In Fig.~\ref{fig3}\textbf{b}(i), a small \(20\,\mathrm{nm}\) displacement from the membrane center lifts this exact cancellation and gives the even-\(m\) modes a finite optical response. The optical response also depends on the probe size: as the beam radius $w_{\rm R}$ increases, the measurement averages
over more regions of opposite displacement, reducing the visibility even when the mechanical amplitude remains appreciable. Importantly, preparation and optical readout use different spatial weightings of the same mechanical mode, so their overlaps need not coincide: a mode can be efficiently driven yet remain weakly visible to a Gaussian probe (Supplementary Secs. S3.3, S3.4 and S5.1).

By contrast, the electron response is governed by coherent phase
accumulation along its trajectory rather than by the intensity-weighted
spatial averaging of the optical measurement. Each position along the trajectory contributes to the coupling with a phase determined by the electron motion relative to the mechanically generated optical near field. At $\Delta_m=0$, these contributions add coherently rather than cancel,
allowing finite coupling to both even- and odd-$m$ modes
(Fig.~\ref{fig3}\textbf{a}).

Figure~\ref{fig3}\textbf{b}(ii) shows that the coupling maximum shifts
to lower kinetic energy as $m$ increases, because the larger $K_m$
requires a lower phase-matched velocity. Tuning the electron energy therefore preferentially selects the spatial Fourier components of the mechanically generated field sampled along the trajectory, providing mode-selective readout without tighter optical focusing. This selectivity is not perfectly exclusive. Thus, neighboring mechanical modes remain partly coupled (Supplementary Fig. S2).

\begin{figure}[!t]
\centering
\centerline{\includegraphics[width=\textwidth]{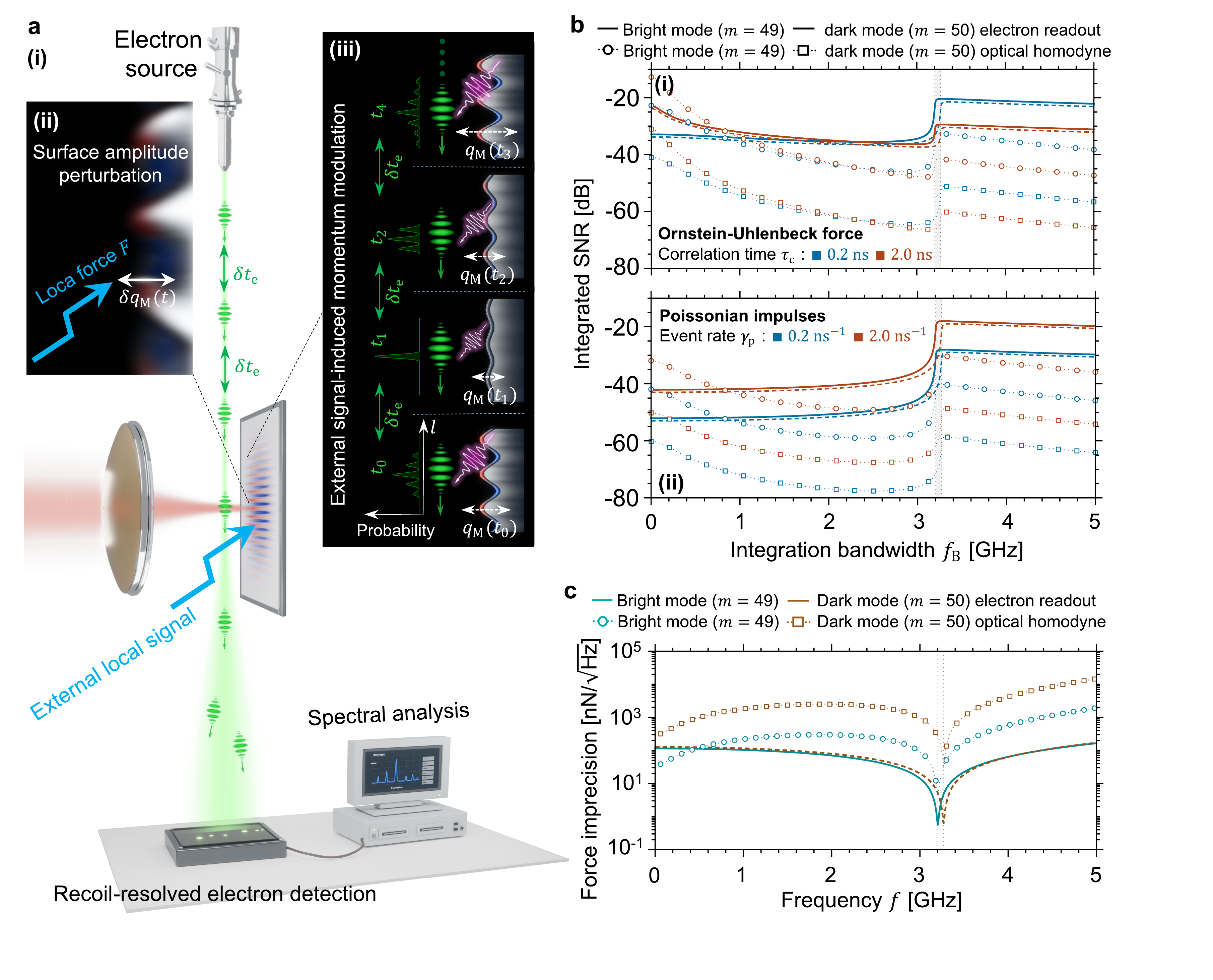}} \caption{\textbf{Free-electron readout of localized mechanical perturbations.}
\textbf{a}, Proposed sensing geometry.
\textbf{(i)}, Successive electrons probe an optically illuminated membrane,
with $\delta t_{\rm e}$ denoting the interval between arrivals.
\textbf{(ii)}, A localized external force perturbs the amplitude of a
transverse mechanical mode and thereby modifies the mechanically generated
optical near field.
\textbf{(iii)}, The resulting change in electron--field coupling modulates
the electron momentum distribution, which is measured using energy-resolved electron detection. Momentum-dispersive electron optics required to resolve the electron recoil orders are omitted for clarity. Schematics are not to scale.
\textbf{b}, Integrated SNR for an Ornstein--Uhlenbeck force with
correlation times $\tau_{\rm c}=0.2,\,2.0\,{\rm ns}$ (\textbf{i}) and
mean-subtracted Poisson impulses with event rates
$\gamma_{\rm P}=0.2,\,2.0\,{\rm ns}^{-1}$ (\textbf{ii}).
The SNR increases strongly as the integration bandwidth reaches the
mechanical resonance, with electron readout retaining a substantially
larger response than Gaussian homodyne detection for the optically
near-dark $m=50$ mode. Blue and orange denote the two signal
parameters. Solid and dashed curves show electron readout for
$m=49$ and $m=50$, respectively; dotted curves with open circles and
squares show the corresponding optical readout.
\textbf{c}, Force-referred readout imprecision for the same modes and
readout schemes. The imprecision reaches a minimum near each mechanical
resonance, where the mechanical response is enhanced. Electron readout
retains similar sensitivity for the two modes, whereas optical
cancellation strongly degrades the $m=50$ response. Teal and brown
denote $m=49$ and $m=50$, respectively, and vertical dotted lines mark
their resonances. The modes are evaluated separately as single-mode
benchmarks, with the electron energy phase matched to the selected mode
and equal detected electron and optical probe rates of
$10^9\,{\rm s}^{-1}$.}
\label{fig4}
\end{figure}

This contrast becomes useful for sensing an external signal represented
by a localized force (Fig.~\ref{fig4}\textbf{a}). A force applied near
an antinode can efficiently drive the mechanical mode even when its
Gaussian optical response remains suppressed \cite{Esmenda2021}.
For a Gaussian force profile of rms spatial width $\sigma_{\rm F}$
centered at $x_{\rm s}$ and $y=L/2$, the driving strength of a given
mode is determined by the spatial overlap between the force profile and
the mechanical mode shape. We describe this by the dimensionless
force--mode coupling factor
$H_m=\exp[-K_m^2\sigma_{\rm F}^2/2]\sin(K_mx_{\rm s})$.
The force spectrum $\tilde F(f)$, where $f$ is the Fourier frequency,
produces a force-induced displacement perturbation $\delta\tilde q_m(f)=H_m\chi_m(f)\tilde F(f)$ of mechanical mode $m$, where
$\chi_m$ is the mechanical susceptibility that converts modal force into displacement \cite{Aspelmeyer2014}. The resulting mechanical motion then encodes the external signal in the electron recoil populations.

We compare electron readout with Gaussian optical homodyne detection
for two neighboring mechanical modes: the optically brighter $m=49$ and the near-dark $m=50$ modes. Each mode is considered separately at equal detected electron and photon rates, with the electron energy set for phase matching. For a given mode and force, both measurements probe the same force-induced mechanical displacement. Therefore, differences in sensitivity arise from how each measurement responds to the same mechanical displacement (Methods).

To test the sensing response to time-dependent external forces,
Fig.~\ref{fig4}\textbf{b} considers continuously correlated
Ornstein--Uhlenbeck fluctuations and randomly arriving, mean-subtracted
Poisson impulses, representing correlated force noise and discrete force
events, respectively. Because these signals are broadband, their
detectability is not captured by the response at a single frequency.
We therefore use an integrated signal-to-noise ratio (SNR), which
measures the accumulated signal relative to noise across the integration bandwidth:
\begin{equation}
\mathrm{SNR}_r(f_{\rm B})
=
\frac{
\displaystyle\int_0^{f_{\rm B}}
|H_m\chi_m(f)|^2S_{\rm F}(f)\,\mathrm{d}f
}{
\displaystyle\int_0^{f_{\rm B}}
\left[S_r^{\rm imp}(f)+S_m^{\rm th}(f)\right]\,\mathrm{d}f
}.
\label{eq:main_integrated_snr}
\end{equation}
Here $r$ denotes electron or optical detection and $f_{\rm B}$ is the upper limit of the integration bandwidth. The spectra $S_{\rm F}$,
$S_r^{\rm imp}$, and $S_m^{\rm th}$ describe the applied force,
readout imprecision, and thermal mechanical noise, respectively
\cite{Clerk2010,Aspelmeyer2014}. Varying $f_{\rm B}$ then shows how the overall sensing performance changes as a wider frequency range is included.

The force spectrum determines how much signal lies near the mechanical resonance. For Ornstein--Uhlenbeck forcing, the
shorter correlation time places more spectral weight near resonance
and therefore produces a larger SNR, whereas for Poisson forcing a
higher event rate increases the signal strength. In both cases, the
integrated SNR rises sharply as the integration bandwidth reaches the
mechanical resonances near $3.2\,\mathrm{GHz}$, where the mechanical
susceptibility $\chi_m$ enhances the displacement response. Beyond resonance,
additional bandwidth contributes little signal while continuing to
accumulate noise, so the SNR can decrease.

For the adopted coupling calibration and equal detected-probe rates,
electron detection at $f_{\rm B}=5\,{\rm GHz}$ exceeds Gaussian
homodyne detection by approximately $16\,{\rm dB}$ for $m=49$ and
$34\,{\rm dB}$ for $m=50$. The larger difference for the near-dark
$m=50$ mode results from its suppressed optical overlap, while the
phase-matched electron coupling remains appreciable. The absolute SNR
remains below unity under the prescribed signal calibration, so these
values quantify the relative performance of the two detection schemes.

Force-referred imprecision gives a comparison independent of the
chosen force spectrum (Fig.~\ref{fig4}\textbf{c}), by expressing
displacement-readout noise as an equivalent applied force,
\begin{equation}
\sqrt{S_{{\rm F},r}^{\rm imp}(f)}
=
\frac{\sqrt{S_r^{\rm imp}(f)}}{|H_m\chi_m(f)|}.
\label{eq:main_force_imprecision}
\end{equation}
The spatial force profile remains part of this comparison through
$H_m$. The resonant minima arise because the mechanical susceptibility is enhanced near resonance, reducing the force strength required to produce a displacement comparable to the readout noise and therefore improving the force sensitivity. The two modes retain comparable phase-matched electron coupling, whereas optical cancellation strongly suppresses the readout sensitivity of the $m=50$ mode. At their respective resonances,
electron imprecision is approximately $0.57$ and
$0.64\,\mathrm{nN\,Hz^{-1/2}}$, compared with $4.0$ and
$34\,\mathrm{nN\,Hz^{-1/2}}$ for homodyne. Optical readout can be favorable for the brighter mode at low frequencies, but cavity filtering weakens its GHz response. The comparison therefore depends on both the spatial measurement profile and the frequency range over which the motion is detected. This frequency-domain analysis characterizes the sensitivity to harmonic mechanical motion and does not by itself imply reconstruction of a GHz force waveform from individual electron measurements.

\section*{Experimental roadmap and operating regime}

An open Fabry--P\'erot cavity provides a plausible geometry combining
optical confinement with lateral electron access. Reflective
Si$_3$N$_4$ membranes and open micro- and fiber cavities establish the
main ingredients of this geometry
\cite{Chen2017,Greuter2014,Tenbrake2024,Rochau2021}. In the reference
configuration, the mechanically generated evanescent field has a decay
length of approximately $0.21\,\mu\mathrm{m}$, much shorter than the
cavity length $L_{\rm c}=10\,\mu\mathrm{m}$, while the optical quality factor
$Q_{\rm opt}=10^6$ corresponds to a finesse near $5\times10^4$.
Because the electron samples this local boundary field, the sideband
need not itself be resonantly stored as a propagating cavity mode
\cite{CheungLaw2011,Aspelmeyer2014}. The absolute moving-boundary
conversion remains material-dependent and is therefore treated as a
prescribed calibration.

The mechanical requirement is a high-order GHz mode with the spatial
period required for phase matching. For $L=20\,\mu\mathrm{m}$, the
assigned $(50,1)$ mode at $3.27\,\mathrm{GHz}$ corresponds to a
transverse wave speed near $2.6\,\mathrm{km\,s^{-1}}$ and may require
an engineered membrane or phononic structure; GHz radiation-pressure-driven motion provides a relevant precedent \cite{Krause2015}.
The mode can be prepared with non-collinear optical beams, which for
the reference wavelength require incidence near $\pm38.7^\circ$ from
the normal, or by piezoelectric or acoustic actuation
\cite{Iyer2024,Jiang2020}. Independent preparation avoids the strong
filtering of $3.27\,\mathrm{GHz}$ intensity modulation by the
$300\,\mathrm{MHz}$ interrogation-cavity linewidth, so the
$8\,\mathrm{mW}$ interrogation power should not be identified with the
preparation power.

The preparation and sensing benchmarks use the mechanical quality factor $Q_{\rm M}=10^4$ and $200$,
respectively, and therefore represent different operating conditions.
The reference displacement remains within the first-order
moving-boundary regime, while the approximately $80\,\mathrm{fs}$
electron crossing time is much shorter than the $0.31\,\mathrm{ns}$
mechanical period, supporting a frozen mechanical configuration during
each passage. Neither ground-state preparation nor single-photon
optomechanical strong coupling is required; residual lower-sideband coupling and its suppression are discussed in Supplementary Secs.~S4.6--S4.7. Modes $m=49$ and $50$ require approximately $374$ and $341\,\mathrm{keV}$ electron energies at
a $70\,\mathrm{nm}$ central clearance. Beam divergence, membrane
flatness, charging, alignment, electron energy resolution, and
spectrometer throughput will therefore constrain the accessible
interaction.

Localized forcing is compatible with established nanomechanical
actuation and force-microscopy techniques
\cite{Vicary2011,Rossi2017,Antognozzi2016,Esmenda2021}. These
precedents support spatially localized forcing; the force strengths
used here remain prescribed benchmarks. The reference width
$\sigma_{\rm F}=20\,{\rm nm}$ should be regarded as a benchmark for
efficient coupling to the high-spatial-frequency mode rather than a
requirement of the electron readout. A practical first experiment
could measure recoil modulation under a calibrated mechanical drive
before introducing an unknown signal. A device-specific sensitivity
analysis would additionally need to include technical noise, modal
cross-talk, heating, and optical or electron backaction.

\section*{Significance and outlook}

The present framework uses mechanical motion to control the outgoing electron state through an externally driven optical field. This differs from surface-wave-assisted Smith--Purcell radiation, in which a moving interface scatters the electron's electromagnetic field into outgoing light~\cite{Saharian2021}. Here, the vibrating boundary instead scatters an optical carrier into an evanescent field that coherently modulates the electron. The mechanically scattered field therefore serves as a controllable drive for electron-state manipulation rather than as the radiative output. At a fixed mechanical mode, adjusting the vibration amplitude redistributes the electron population among recoil orders without shifting the phase-matched energy. External forces that perturb the vibration also modify these populations, allowing localized-force signals to be read out from the electron recoil spectrum.

Building on this recoil control, mechanical actuation offers strategies that do not rely solely on reshaping the incident optical field, complementing established optical manipulation of electron wavefunctions~\cite{bibKD6,bibPINEM24}. In suitable devices, electrical or acoustic excitation could tune the coupling while the incident optical carrier remains fixed~\cite{Jiang2020}. A prepared vibration can also persist after actuation ends~\cite{Aspelmeyer2014}; under continued optical illumination, its decaying amplitude could then modulate successive electron interactions. Selecting or combining mechanical modes could further reconfigure the spatial spectrum of the coupling field without changing the illumination pattern, subject to mode-dependent phase matching. Beyond amplitude control, a phase-coherent reference interaction addressing the same electron transitions could convert the retained mechanical phase into recoil-population contrast~\cite{Ropers2016,bibPINEM13}. These extensions lie beyond the present calculations and would require accounting for mechanical bandwidth, thermal fluctuations, electron timing, and relative-phase stability.

The demonstrated mode selectivity also motivates spatially resolved force spectroscopy. High-order mechanical motion that is weakly visible to a Gaussian optical probe can remain accessible through phase-matched electron coupling, complementing localized optomechanical readout~\cite{Roxworthy2018,Roxworthy2016}. Because a localized force drives mechanical modes with weights determined by its spatial profile~\cite{Esmenda2021}, measurements at different electron energies and trajectories could help constrain the position and extent of an unknown perturbation. A joint analysis of several modes would need to establish which spatial features can be recovered in the presence of modal cross-talk and measurement noise. In such a measurement, the mechanical resonator would provide spatial and frequency filtering, while electron recoil would read out the resulting modal response.

Beyond prescribed classical motion, a quantum treatment of the mechanical mode and optical mediator could test whether phonon states with the same coherent displacement but different fluctuations produce distinguishable electron spectra. Mechanical-state tomography~\cite{Vanner2013}, the transfer of photon statistics to electrons~\cite{bibPINEM18}, and electron-conditioned optical state preparation~\cite{bibFree2} provide relevant precedents. Further questions include the generation of electron--phonon correlations and the possibility of conditioning mechanical states on electron detection. Such capabilities would require explicit treatment of decoherence, loss, and backaction, while any proposed quantum-sensing advantage would additionally require a resource-matched comparison with optical detection.

\section*{Methods}

\subsection*{Mechanical modes and mechanically generated optical fields}

We model the vibrating element as a square membrane of side length
$L$, with its equilibrium surface at $z=L_{\rm c}$ and displacement
normal to the membrane plane. In the fixed-edge, tension-dominated
limit, where the restoring force is governed primarily by in-plane
tensile stress and flexural rigidity is neglected
\cite{Joeckel2011}, the peak-normalized mechanical eigenmodes are
\begin{equation}
\psi_{mn}(x,y)
=
\sin\!\left(\frac{m\pi x}{L}\right)
\sin\!\left(\frac{n\pi y}{L}\right),
\qquad
\Omega_{mn}
=
\frac{\pi v_{\rm M}}{L}
\sqrt{m^2+n^2},
\label{eq:methods_modes}
\end{equation}
with $\max|\psi_{mn}|=1$.
We focus on the $(m,1)$ modes and write
$\psi_m\equiv\psi_{m1}$ and $K_m=m\pi/L$.
Here \(v_{\rm M}\) is the transverse wave speed of the tension-dominated membrane. We use the dispersion as an effective modal model, calibrating \(v_{\rm M}\) to the reference \(m=50\) frequency and applying the same relation to neighboring modes.

The displacement operator of mode $m$ is the harmonic-oscillator
quadrature weighted by the mechanical eigenmode profile,
\begin{equation}
\hat Q_m(x,y)
=
x_{{\rm zpf},m}\psi_m(x,y)
\left(\hat b_m+\hat b_m^\dagger\right),
\qquad
x_{{\rm zpf},m}
=
\sqrt{\frac{\hbar}{2M_{\rm eff}\Omega_m}},
\end{equation}
where $\hat b_m$ ($\hat b_m^\dagger$) annihilates (creates) a phonon in
mode $m$, $M_{\rm eff}$ is its effective modal mass, and
$x_{{\rm zpf},m}$ is the zero-point displacement.

The moving dielectric boundary couples this mechanical mode to the
optical field. To first order in the displacement, the corresponding
photon--phonon interaction can be written as
\cite{Sipe2016,CheungLaw2011}
\begin{equation}
\hat H_{\rm MB}
=
\hbar
\sum_{\lambda,\lambda',m}
G_{\lambda'\lambda m}
\hat a_{\lambda'}^\dagger
\hat a_\lambda
\left(\hat b_m+\hat b_m^\dagger\right),
\end{equation}
where $\hat a_\lambda$ ($\hat a_\lambda^\dagger$) annihilates (creates) a photon in optical mode $\lambda$, and $G_{\lambda'\lambda m}$ contains the electromagnetic moving-boundary overlap and the mechanical
zero-point displacement. This interaction describes optical scattering
accompanied by phonon exchange, providing the microscopic origin of the mechanically generated sidebands.

For the coherently driven mechanical motion considered here, the coherent displacement is much larger than the zero-point and thermal mechanical fluctuations. We also assume negligible electron-induced backaction during each passage. In this large-coherent-amplitude,
weak-backaction regime, we replace the mechanical displacement operator by its expectation value (Supplementary Secs.~S1.4 and S8.4),
\[
\langle\hat Q_m(x,y,t)\rangle
=
q_{\rm M}\psi_m(x,y)\cos(\Omega_m t+\phi_m),
\qquad
q_{\rm M}=2x_{{\rm zpf},m}|\alpha_m|,
\]
where $\alpha_m=|\alpha_m|e^{-i\phi_m}$. With the peak normalization of $\psi_m$, $q_{\rm M}$ is the displacement
amplitude at a mechanical antinode. We therefore treat the membrane as
a prescribed time-dependent dielectric boundary and obtain the
resulting optical field using first-order moving-boundary perturbation
theory \cite{Johnson2002}. A small displacement $Q_m(x,y,t)$ changes
both the position and local orientation of the interface, generating
an optical-field correction linear in the displacement and its
in-plane gradient. We retain these first-order terms for
$k_{\rm L}q_{\rm M}\ll1$ and $K_mq_{\rm M}\ll1$, where $k_{\rm L}=\omega_{\rm L}/c$, while neglecting bulk photoelasticity and surface-velocity corrections.

The resulting standing mechanical pattern contains spatial components
$\pm K_m$ and temporal components $\pm\Omega_m$. Illuminating it with
a carrier at $\omega_{\rm L}$ therefore generates optical components
at $\omega_{ms}=\omega_{\rm L}+s\Omega_m$ and tangential wavevectors
$\pm K_m$, with $s=\pm1$. In vacuum, a component becomes evanescent
when $K_m>\omega_{ms}/c$, with decay constant
\begin{equation}
\kappa_{ms}
=
\sqrt{
K_m^2-\frac{\omega_{ms}^2}{c^2}
}.
\label{eq:methods_decay}
\end{equation}
The corresponding field decays as
$e^{-\kappa_{ms}(L_{\rm c}-z)}$ away from the membrane. In the
calculations below we retain the upper temporal sideband,
$s=+1$, and write $\kappa_m=\kappa_{m,+}$.

A convenient local form for the positive-frequency vector potential is
\begin{equation}
\mathbf A_{ms}^{(+)}(\mathbf r,t)
=
A_0\xi_{ms}u(x,y)
e^{-\kappa_{ms}(L_{\rm c}-z)}
e^{-i\omega_{ms}t-is\phi_m}
\sum_{\sigma=\pm1}
\frac{\sigma}{2i}
\boldsymbol{\epsilon}_{m\sigma s}
e^{i\sigma K_mx}.
\label{eq:methods_sideband_field}
\end{equation}
Here $u(x,y)$ is the slowly varying illumination envelope and
$\boldsymbol{\epsilon}_{m\sigma s}$ is the TM polarization vector of
the corresponding evanescent Fourier component. The dimensionless
sideband amplitude $\xi_{ms}$ is linear in the mechanical displacement
for small motion, $\xi_{ms}\simeq\eta_{ms}k_{\rm L}q_{\rm M}$.
The sideband conversion coefficient $\eta_{ms}$ depends on the dielectric contrast, electromagnetic mode profiles,
polarizations, and cavity geometry
\cite{Johnson2002,CheungLaw2011,Sipe2016}. Because its absolute value
is device-dependent, we absorb its common magnitude into the calibrated
reference coupling $g_{\rm ref}$ rather than assigning it independently.
The carrier amplitude scales as
$A_0\propto\sqrt{P_{\rm in}Q_{\rm opt}}$ for fixed geometry and
port-coupling fraction.

The standing mechanical motion generates both upper and lower optical sidebands. In what follows, we focus on the upper optical sideband at $\omega_{\rm L}+\Omega_m$; the lower-sideband contribution can be included analogously when it is not experimentally suppressed.

The phonon-assisted electron interaction is then described, in the
coherent-field limit, by the coupling of the electron to this
mechanically generated optical field. Neglecting the small change in
electron velocity during a single interaction event, the
interaction-picture Hamiltonian is
\begin{equation}
\hat H_{m}^{I}(t)
=
e\,\mathbf v\cdot\mathbf A^{(+)}_{m,+}
\!\left(\hat{\mathbf r}+\mathbf vt,t\right)+\mathrm{H.c.},
\label{eq:methods_electron_interaction}
\end{equation}
where $-e$ is the electron charge,
$\mathbf v$ is its central velocity, and
$\mathbf A_{m,+}^{(+)}$ is the positive-frequency vector potential
of the mechanically generated upper sideband defined above.
The translation
$\hat{\mathbf r}\rightarrow\hat{\mathbf r}+\mathbf vt$
describes free electron propagation in the interaction picture.
This prescribed-trajectory approximation retains the quantum
scattering of the electron while neglecting recoil-induced changes
of its velocity during the interaction \cite{Abajo2010}.

\subsection*{Electron coupling and recoil scattering}

The electron trajectory is taken as
$\mathbf r_{\rm e}(t)=
(x_{\rm c}+v\cos\theta\,t,
y_{\rm c},
L_{\rm c}-d+v\sin\theta\,t)$,
with $x_{\rm c}=y_{\rm c}=L/2$.
The angle $\theta$ is measured from the membrane plane and $d$ is the
electron--membrane separation at the interaction center.

The complex coupling amplitude is obtained by integrating the
upper-sideband field along this trajectory,
\begin{equation}
\tilde g_m
=
\frac{e}{\hbar}
\int dt\,
\mathbf v\cdot
\mathbf A_{m,+}^{(+)}
[\mathbf r_{\rm e}(t),t],
\qquad
g_m=2|\tilde g_m|.
\label{eq:methods_coupling_integral}
\end{equation}
This trajectory integral is the basis of the finite-membrane
calculations used for the mode-resolved and sensing results. It retains
both spatial harmonics $\pm K_m$, their polarization projections, the
evanescent decay, and the change in electron--surface separation along
the tilted trajectory.

For intuition, the illumination along the electron direction is
approximated by a Gaussian envelope of radius $w_{\rm e}$. The
corresponding effective interaction time is
$\tau_{\rm int}=w_{\rm e}/(v\cos\theta)$. Using the phase mismatch
$\Delta_m$ defined in Eq.~\eqref{eq:phase_matching}, the analytical
Gaussian-envelope result reduces to
\begin{equation}
g_m
\simeq
g_{\rm ref}
\frac{q_{\rm M}}{q_{\rm ref}}
e^{-\kappa_m d}
\exp\!\left[
-\frac{
v^2\Delta_m^2\tau_{\rm int}^2
}{4}
\right].
\label{eq:methods_simple_coupling}
\end{equation}
The three factors describe mechanical-amplitude control, evanescent
attenuation, and finite-time phase-matching selectivity. The exact
finite-membrane calculation evaluates
Eq.~\eqref{eq:methods_coupling_integral} numerically rather than using
this Gaussian approximation.

We use $q_{\rm ref}=2.5\,{\rm nm}$ and define
$g_{\rm ref}=5.72$ as the calibrated phase-matched coupling scale at
the membrane surface. At the operating separation $d=70\,{\rm nm}$,
the $m=50$ mode then gives $g_{50}=4.11$ at its phase-matched
electron energy. This calibration absorbs the carrier amplitude \(A_0\), the reference moving-boundary conversion coefficient \(\eta_{50,+}\), and the corresponding polarization and geometrical factors.

The interaction acts as a phase modulation of the electron wavefunction.
For a slowly varying envelope,
\begin{equation}
\hat S_m
=
\exp\!\left[
-i g_m
\cos\!\left(
\frac{\omega_{\rm L}+\Omega_m}{v\cos\theta}(\hat x-x_{\rm c})+\Phi_m
\right)
\right].
\label{eq:methods_scattering_operator}
\end{equation}
Expanding
Eq.~\eqref{eq:methods_scattering_operator} in recoil orders gives the
standard Bessel populations \cite{Abajo2010,Feist2015},
\begin{equation}
P_\ell
=
J_\ell^2(g_m),
\qquad
\sum_\ell P_\ell=1,
\qquad
\langle\ell^2\rangle=\frac{g_m^2}{2}.
\label{eq:methods_recoil_probabilities}
\end{equation}
At phase matching ($\Delta_m=0$),
$(\omega_{\rm L}+\Omega_m)/(v\cos\theta)=K_m$,
so adjacent recoil orders are separated by $\hbar K_m$
in tangential momentum.
The mechanical phase is contained in $\Phi_m$ and therefore affects the complex scattering amplitudes, but not the populations of resolved,
isolated recoil orders. For Fig.~\ref{fig1}\textbf{b} and
Fig.~\ref{fig3}\textbf{b}(ii), the displayed coupling is normalized
globally as
$g_{\rm N}=g_m^2/\max_{\mathcal D}g_m^2$ over the complete plotted
domain $\mathcal D$; all sensing calculations use the absolute
coupling.

\subsection*{Spatial overlap in preparation and optical readout}

Mechanical preparation and Gaussian optical readout are described by
the normalized spatial overlap $\mathcal O_m[\mathcal W]$ defined in
Eq.~\eqref{eq:main_mode_overlap}, with the weighting
$\mathcal W(x,y)$ determined by the optical process. For preparation,
we evaluate the full two-dimensional overlap using the focused and
two-beam profiles $\mathcal W_{\rm F}$ and $\mathcal W_{\rm B}$
introduced in the main text, retaining the finite mechanical-mode
profile along both $x$ and $y$.

At mechanical resonance, the ideal-reflector calibration gives the
peak displacement
\begin{equation}
q_{\rm M}
=
\frac{
2Q_{\rm M}
\mathcal O_m[\mathcal W]
P_{\rm AC}
}{
cM_{\rm eff}\Omega_m^2
},
\label{eq:methods_preparation}
\end{equation}
where $P_{\rm AC}$ and $Q_{\rm M}$ are the local optical modulation-power amplitude at the membrane and the mechanical quality factor, respectively. The quantity $2P_{\rm AC}/c$ sets the radiation-pressure force scale for an ideal reflector; the force amplitude driving the selected mode
is $2P_{\rm AC}\mathcal{O}_m[\mathcal{W}]/c$ (Supplementary Sec.~S5.2). This relation is used for Fig.~\ref{fig2}\textbf{c}; it refers to the power delivered
locally to the membrane rather than the external laser power before cavity coupling or other losses.

For Gaussian interferometric readout, we take
$\mathcal W_{\rm R}=I_{\rm G}$, with
\begin{equation}
I_{\rm G}(x,y)
\propto
\exp\!\left[
-\frac{2[(x-x_{\rm R})^2+(y-y_{\rm R})^2]}{w_{\rm R}^2}
\right],
\label{eq:intensityprofile}
\end{equation}
where $(x_{\rm R},y_{\rm R})$ denotes the center of the readout beam
and $w_{\rm R}$ is its $1/e^2$ intensity radius.
The dimensionless overlap $\mathcal{O}_m[\mathcal{W}_{\rm R}]$ gives the optical visibility
of the mechanical mode \cite{Haghighi2018}.
For a signed modal displacement $q_m(t)$, the effective displacement sampled by the receiver has magnitude $\mathcal{O}_m[\mathcal{W}_{\rm R}]|q_m(t)|$. A centered beam suppresses modes whose
positive and negative displacements cancel under this weighting,
whereas a finite lateral offset weakly lifts the symmetry cancellation.
The finite-area integrals used in Fig.~\ref{fig3} are evaluated
numerically; the corresponding analytical Gaussian approximation is given in the Supplementary Information.

For a small signed modal displacement $q_m(t)$, the magnitude of the resulting cavity-frequency shift is
\begin{equation}
|\delta\omega_{\rm c}|
\simeq
\frac{\omega_{\rm L}}{L_{\rm c}}
\mathcal O_m[\mathcal W_{\rm R}]
|q_m|,
\label{eq:cavity_frequency_shift}
\end{equation}
which provides the spatial transduction factor used in the Gaussian homodyne benchmark \cite{Aspelmeyer2014}.

\subsection*{Localized-force sensing and readout imprecision}

We represent the external signal by a force localized near
$y=L/2$. In modal analysis, the force driving a mechanical mode is
obtained by projecting the spatial force distribution onto the
corresponding mode shape \cite{Esmenda2021}. Writing the force profile as $F(t)h(x-x_{\rm s})$, with $\int h(x-x_{\rm s})\,dx=1$, the dimensionless projection onto mode $m$ is
\begin{equation}
H_m(x_{\rm s})
=
\int
h(x-x_{\rm s})
\sin(K_mx)\,dx .
\label{eq:methods_force_overlap}
\end{equation}
For a Gaussian force profile with rms spatial width $\sigma_{\rm F}$,
this becomes
$H_m=
\exp(-K_m^2\sigma_{\rm F}^2/2)\sin(K_mx_{\rm s})$
away from the membrane edges. Thus $H_m$ contains the spatial
selection of the force, whereas the mechanical susceptibility contains
its frequency response.

The corresponding force-induced modal displacement is
\begin{equation}
\delta\tilde q_m(f)
=
H_m\chi_m(f)\tilde F(f),
\qquad
\chi_m(f)
=
\frac{1}{
M_{\rm eff}
\left[
\Omega_m^2-(2\pi f)^2
-i(2\pi f)\Omega_m/Q_{\rm M}
\right]
}.
\label{eq:methods_mechanical_response}
\end{equation}
The calculations in Fig.~\ref{fig4} treat $m=49$ and $50$ as
separate single-mode benchmarks. Electron-energy tuning is not assumed
to eliminate neighboring-mode responses; such contributions would
need to be identified in a multimode measurement.

For electron readout, we use the information contained in the complete
recoil distribution $P_\ell=J_\ell^2(g_m)$ rather than a single
recoil order. The corresponding Fisher information for the coupling
amplitude is $\mathcal I_g=2$, giving the Fisher-information rate
\begin{equation}
\dot{\mathcal I}_{\rm e}(f)
=
2R_{\rm e}
\left|
\frac{\partial g_m}{\partial q_{\rm M}}
\right|^2,
\label{eq:methods_electron_FI}
\end{equation}
where $R_{\rm e}$ is the detected electron rate. In the linear
moving-boundary regime,
$\partial g_m/\partial q_{\rm M}=g_m/q_{\rm M}$.
For a displacement component at Fourier frequency $f$, the electron
kernel is evaluated at the corresponding upper optical sideband
$\omega_{\rm L}+2\pi f$. The derivation of
$\mathcal I_g=2$ and the treatment of the zero-amplitude limit are
given in the Supplementary Information.

For comparison, the optical phase response to displacement of a resonantly
driven Gaussian homodyne receiver is
\begin{equation}
\mathcal R_m^{\rm opt}(f)
=
\mathcal O_m[\mathcal W_{\rm R}]
\frac{\omega_{\rm L}/L_{\rm c}}
{\kappa_{\rm c}/2-i2\pi f}
\frac{1}{1-if/f_{\rm det}},
\qquad
\dot{\mathcal I}_{\rm o}(f)
=
2R_{\rm ph}
|\mathcal R_m^{\rm opt}(f)|^2 ,
\label{eq:methods_optical_FI}
\end{equation}
Here, $\kappa_{\rm c}=\omega_{\rm L}/Q_{\rm opt}$, $R_{\mathrm{ph}}$
is the detected photon rate, and $f_{\mathrm{det}}$ parameterizes
the effective detector response. The benchmark applies this response
to the signal while retaining a fixed normalized output-noise level.
It therefore differs from a noiseless electronic filter acting
identically on signal and shot noise, as discussed in
Supplementary Sec.~S6.5. We assume ideal detection of the appropriate
optical quadrature \cite{Aspelmeyer2014}.

Using the one-sided displacement-noise convention specified in the
Supplementary Information, the displacement-referred readout
imprecision is
$S_r^{\rm imp}(f)=1/\dot{\mathcal I}_r(f)$ for
$r\in\{{\rm e,o}\}$. The force-referred imprecision then follows from
Eq.~\eqref{eq:main_force_imprecision}. Independent classical thermal
motion contributes
\begin{equation}
S_m^{\rm th}(f)
=
4k_{\rm B}T
M_{\rm eff}
\frac{\Omega_m}{Q_{\rm M}}
|\chi_m(f)|^2,
\label{eq:methods_thermal_noise}
\end{equation}
where all spectral densities are one-sided and expressed per hertz
\cite{Clerk2010}. The mechanical displacement signal spectrum is
$S_m^{\rm sig}(f)=|H_m\chi_m(f)|^2S_{\rm F}(f)$, and the integrated
SNR is then evaluated using Eq.~\eqref{eq:main_integrated_snr}.

Figure~\ref{fig4} compares electron and optical readout at equal
detected probe rates,
$R_{\rm e}=R_{\rm ph}=10^9\,{\rm s}^{-1}$, using
$Q_{\rm M}=200$. Figure~\ref{fig2} instead uses
$Q_{\rm M}=10^4$ for the mechanical-preparation benchmark. Equal detected rates do not imply equal total resources or use of
all available optical photons. The optical field used for the electron
interaction and the Gaussian readout beam are assigned separate
spatial profiles and are not assumed to be the same cavity mode
(Supplementary Sec.~S3.4).
The definitions and normalization of the Ornstein--Uhlenbeck and
Poisson force spectra, the complete parameter set, stochastic-signal
calibration and numerical implementation are provided in the
Supplementary Information.

\section*{Data availability}

The numerical data supporting the findings of this study are available
from the corresponding author upon reasonable request.

\section*{Code availability}

The custom numerical code used to generate the results and figures in
this study is available from the corresponding author upon reasonable
request.

\bibliography{sn-bibliography}

@article{bibOpt0,
  author        = "Barzanjeh, S. and Xuereb, A. and Gr{\"o}blacher, S. and Paternostro, M. and Regal, C. A. and Weig, E. M.",
  title         = "Optomechanics for quantum technologies",
  journal       = "Nat. {P}hys.",
  volume        = "18",
  number        = "1",
  pages         = "15--24",
  year          = "2022",
  doi           = "10.1038/s41567-021-01402-0"
}

@article{bibOpt2,
  author        = "Kim, P. H. and Hauer, B. D. and Doolin, C. and Souris, F. and Davis, J. P.",
  title         = "Approaching the standard quantum limit of mechanical torque sensing",
  journal       = "Nat. {C}ommun.",
  volume        = "7",
  number        = "1",
  pages         = "13165",
  year          = "2016",
  doi           = "10.1038/ncomms13165"
}

@article{bibOpt6,
  author        = "Arcizet, O. and Cohadon, P.-F. and Briant, T. and Pinard, M. and Heidmann, A. and Mackowski, J.-M. and Michel, C. and Pinard, L. and Fran{\c{c}}ais, O. and Rousseau, L.",
  title         = "High--sensitivity optical monitoring of a micromechanical resonator with a quantum--limited optomechanical sensor",
  journal       = "Phys. {R}ev. {L}ett.",
  volume        = "97",
  number        = "13",
  pages         = "133601",
  year          = "2006",
  doi           = "10.1103/PhysRevLett.97.133601"
}

@article{Aspelmeyer2014,
  author        = "Aspelmeyer, M. and Kippenberg, T. J. and Marquardt, F.",
  title         = "Cavity optomechanics",
  journal       = "Rev. {M}od. {P}hys.",
  volume        = "86",
  number        = "4",
  pages         = "1391--1452",
  year          = "2014",
  doi           = "10.1103/RevModPhys.86.1391"
}

@article{bibOpt8,
  author        = "Stannigel, K. and Rabl, P. and S{\o}rensen, A. S. and Zoller, P. and Lukin, M. D.",
  title         = "Optomechanical transducers for long--distance quantum communication",
  journal       = "Phys. {R}ev. {L}ett.",
  volume        = "105",
  number        = "22",
  pages         = "220501",
  year          = "2010",
  doi           = "10.1103/PhysRevLett.105.220501"
}

@article{bibOpt9,
  author        = "Zhao, H. and Chen, W. D. and Kejriwal, A. and Mirhosseini, M.",
  title         = "Quantum--enabled microwave--to--optical transduction via silicon nanomechanics",
  journal       = "Nat. {N}anotechnol.",
  volume        = "20",
  number        = "5",
  pages         = "602--608",
  year          = "2025",
  doi           = "10.1038/s41565-025-01874-8"
}

@article{bibPINEM1,
  author        = "Barwick, B. and Flannigan, D. J. and Zewail, A. H.",
  title         = "Photon--induced near--field electron microscopy",
  journal       = "Nature",
  volume        = "462",
  number        = "7275",
  pages         = "902--906",
  year          = "2009",
  doi           = "10.1038/nature08662"
}

@article{bibPINEM2,
  author        = "Park, S. T. and Lin, M. and Zewail, A. H.",
  title         = "Photon--induced near--field electron microscopy ({P}{I}{N}{E}{M}): theoretical and experimental",
  journal       = "New {J}. {P}hys.",
  volume        = "12",
  pages         = "123028",
  year          = "2010",
  doi           = "10.1088/1367-2630/12/12/123028"
}

@article{Wang2020,
  author        = "Wang, K. and Dahan, R. and Shentcis, M. and Kauffmann, Y. and Ben Hayun, A. and Reinhardt, O. and Tsesses, S. and Kaminer, I.",
  title         = "Coherent interaction between free electrons and a photonic cavity",
  journal       = "Nature",
  volume        = "582",
  number        = "7810",
  pages         = "50--54",
  year          = "2020",
  doi           = "10.1038/s41586-020-2321-x"
}

@article{bibPINEM9,
  author        = "Ruimy, R. and Karnieli, A. and Kaminer, I.",
  title         = "Free--electron quantum optics",
  journal       = "Nat. {P}hys.",
  volume        = "21",
  number        = "2",
  pages         = "193--200",
  year          = "2025",
  doi           = "10.1038/s41567-024-02743-2"
}

@article{bibPINEM11,
  author        = "Kfir, O. and Louren{\c{c}}o-Martins, H. and Storeck, G. and Sivis, M. and Harvey, T. R. and Kippenberg, T. J. and Feist, A. and Ropers, C.",
  title         = "Controlling free electrons with optical whispering--gallery modes",
  journal       = "Nature",
  volume        = "582",
  number        = "7810",
  pages         = "46--49",
  year          = "2020",
  doi           = "10.1038/s41586-020-2320-y"
}

@article{bibPINEM12,
  author        = "Karnieli, A. and Tsesses, S. and Yu, R. and Rivera, N. and Zhao, Z. and Arie, A. and Fan, S. and Kaminer, I.",
  title         = "Quantum sensing of strongly coupled light--matter systems using free electrons",
  journal       = "Sci. {A}dv.",
  volume        = "9",
  number        = "1",
  pages         = "eadd2349",
  year          = "2023",
  doi           = "10.1126/sciadv.add2349"
}

@article{bibPINEM13,
  author        = "Gaida, J. H. and Louren{\c{c}}o-Martins, H. and Sivis, M. and Rittmann, T. and Feist, A. and {Garc{\'i}a de Abajo}, F. J. and Ropers, C.",
  title         = "Attosecond electron microscopy by free--electron homodyne detection",
  journal       = "Nat. {P}hoton.",
  volume        = "18",
  number        = "5",
  pages         = "509--515",
  year          = "2024",
  doi           = "10.1038/s41566-024-01380-8"
}

@article{Feist2015,
  author        = "Feist, A. and Echternkamp, K. E. and Schauss, J. and Yalunin, S. V. and Sch{\"a}fer, S. and Ropers, C.",
  title         = "Quantum coherent optical phase modulation in an ultrafast transmission electron microscope",
  journal       = "Nature",
  volume        = "521",
  number        = "7551",
  pages         = "200--203",
  year          = "2015",
  doi           = "10.1038/nature14463"
}

@article{bibPINEM18,
  author        = "Dahan, R. and Gorlach, A. and Haeusler, U. and Karnieli, A. and Eyal, O. and Yousefi, P. and Segev, M. and Arie, A. and Eisenstein, G. and Hommelhoff, P. and Kaminer, I.",
  title         = "Imprinting the quantum statistics of photons on free electrons",
  journal       = "Science",
  volume        = "373",
  number        = "6561",
  pages         = "{eabj7128}",
  year          = "2021",
  doi           = "10.1126/science.abj7128"
}

@article{bibPINEM19,
  author        = "Liebtrau, M. and Sivis, M. and Feist, A. and Louren{\c{c}}o-Martins, H. and Pazos-P{\'e}rez, N. and Alvarez-Puebla, R. A. and {Garc{\'i}a de Abajo}, F. J. and Polman, A. and Ropers, C.",
  title         = "Spontaneous and stimulated electron--photon interactions in nanoscale plasmonic near fields",
  journal       = "Light {S}ci. {A}ppl.",
  volume        = "10",
  number        = "1",
  pages         = "82",
  year          = "2021",
  doi           = "10.1038/s41377-021-00511-y"
}

@article{Dahan2020,
  author        = "Dahan, R. and Nehemia, S. and Shentcis, M. and Reinhardt, O. and Adiv, Y. and Shi, X. and Be'er, O. and Lynch, M. H. and Kurman, Y. and Wang, K. and Kaminer, I.",
  title         = "Resonant phase--matching between a light wave and a free--electron wavefunction",
  journal       = "Nat. {P}hys.",
  volume        = "16",
  number        = "11",
  pages         = "1123--1131",
  year          = "2020",
  doi           = "10.1038/s41567-020-01042-w"
}

@article{bibPINEM23,
  author        = "Feist, A. and Huang, G. and Arend, G. and Yang, Y. and Henke, J.-W. and Raja, A. S. and Kappert, F. J. and Wang, R. N. and Louren{\c{c}}o-Martins, H. and Qiu, Z. and Liu, J. and Kfir, O. and Kippenberg, T. J. and Ropers, C.",
  title         = "Cavity--mediated electron--photon pairs",
  journal       = "Science",
  volume        = "377",
  number        = "6607",
  pages         = "777--780",
  year          = "2022",
  doi           = "10.1126/science.abo5037"
}

@article{bibPINEM24,
  author        = "Kempers, S. T. and {van Elk}, I. J. M. and {van Leeuwen}, K. A. H. and Luiten, O. J.",
  title         = "Coherent electron phase--space manipulation by combined elastic and inelastic light--electron scattering",
  journal       = "New {J}. {P}hys.",
  volume        = "26",
  pages         = "093026",
  year          = "2024",
  doi           = "10.1088/1367-2630/ad7631"
}

@article{bibKD5,
  author        = "{Garc{\'i}a de Abajo}, F. J. and Kone{\v{c}}n{\'a}, A.",
  title         = "Optical modulation of electron beams in free space",
  journal       = "Phys. {R}ev. {L}ett.",
  volume        = "126",
  number        = "12",
  pages         = "123901",
  year          = "2021",
  doi           = "10.1103/PhysRevLett.126.123901"
}

@article{bibKD6,
  author        = "Chirita Mihaila, M. C. and Weber, P. and Schneller, M. and Grandits, L. and Nimmrichter, S. and Juffmann, T.",
  title         = "Transverse electron--beam shaping with light",
  journal       = "Phys. {R}ev. X",
  volume        = "12",
  number        = "3",
  pages         = "031043",
  year          = "2022",
  doi           = "10.1103/PhysRevX.12.031043"
}

@article{bibKD2,
  author        = "Tsarev, M. and Thurner, J. W. and Baum, P.",
  title         = "Nonlinear--optical quantum control of free--electron matter waves",
  journal       = "Nat. {P}hys.",
  volume        = "19",
  number        = "9",
  pages         = "1350--1354",
  year          = "2023",
  doi           = "10.1038/s41567-023-02092-6"
}

@article{bibKD7,
  author        = "Koz{\'a}k, M. and Eckstein, T. and Sch{\"o}nenberger, N. and Hommelhoff, P.",
  title         = "Inelastic ponderomotive scattering of electrons at a high--intensity optical travelling wave in vacuum",
  journal       = "Nat. {P}hys.",
  volume        = "14",
  number        = "2",
  pages         = "121--125",
  year          = "2018",
  doi           = "10.1038/nphys4282"
}

@article{bibKD8,
  author        = "Ebel, S. and Talebi, N.",
  title         = "Inelastic electron scattering at a single--beam structured light wave",
  journal       = "Commun. {P}hys.",
  volume        = "6",
  number        = "1",
  pages         = "179",
  year          = "2023",
  doi           = "10.1038/s42005-023-01300-2"
}

@article{Clerk2010,
  author        = "Clerk, A. A. and Devoret, M. H. and Girvin, S. M. and Marquardt, F. and Schoelkopf, R. J.",
  title         = "Introduction to quantum noise, measurement, and amplification",
  journal       = "Rev. {M}od. {P}hys.",
  volume        = "82",
  number        = "2",
  pages         = "1155--1208",
  year          = "2010",
  doi           = "10.1103/RevModPhys.82.1155"
}

@article{bibFree1,
  author        = "Ben Hayun, A. and Reinhardt, O. and Nemirovsky, J. and Karnieli, A. and Rivera, N. and Kaminer, I.",
  title         = "Shaping quantum photonic states using free electrons",
  journal       = "Sci. {A}dv.",
  volume        = "7",
  number        = "11",
  pages         = "eabe4270",
  year          = "2021",
  doi           = "10.1126/sciadv.abe4270"
}

@article{bibFree2,
  author        = "Arend, G. and Huang, G. and Feist, A. and Yang, Y. and Henke, J.-W. and Qiu, Z. and Jeng, H. and Raja, A. S. and Haindl, R. and Wang, R. N. and Kippenberg, T. J. and Ropers, C.",
  title         = "Electrons herald non--classical light",
  journal       = "Nat. {P}hys.",
  volume        = "21",
  number        = "11",
  pages         = "1855--1862",
  year          = "2025",
  doi           = "10.1038/s41567-025-03033-1"
}

@article{CheungLaw2011,
  author  = "Cheung, H. K. and Law, C. K.",
  title   = "Nonadiabatic optomechanical {Hamiltonian} of a moving dielectric membrane in a cavity",
  journal = "Phys. Rev. A",
  volume  = "84",
  number  = "2",
  pages   = "023812",
  year    = "2011",
  doi     = "10.1103/PhysRevA.84.023812"
}

@article{Vicary2011,
  author  = "Vicary, J. A. and Ulcinas, A. and Hörber, J. K. H. and Antognozzi, M.",
  title   = "Micro-fabricated mechanical sensors for lateral molecular-force microscopy",
  journal = "Ultramicroscopy",
  volume  = "111",
  number  = "11",
  pages   = "1547--1552",
  year    = "2011",
  doi     = "10.1016/j.ultramic.2011.08.008"
}

@article{Antognozzi2016,
  author  = "Antognozzi, M. and Bermingham, C. R. and Harniman, R. L. and Simpson, S. and Senior, J. and Hayward, R. and Hoerber, H. and Dennis, M. R. and Bekshaev, A. Y. and Bliokh, K. Y. and Nori, F.",
  title   = "Direct measurements of the extraordinary optical momentum and transverse spin-dependent force using a nano-cantilever",
  journal = "Nat. {P}hys.",
  volume  = "12",
  number  = "8",
  pages   = "731--735",
  year    = "2016",
  doi     = "10.1038/nphys3732"
}

@article{Sipe2016,
  author  = "Sipe, J. E. and Steel, M. J.",
  title   = "A {Hamiltonian} treatment of stimulated {Brillouin} scattering in nanoscale integrated waveguides",
  journal = "New J. Phys.",
  volume  = "18",
  number  = "4",
  pages   = "045004",
  year    = "2016",
  doi     = "10.1088/1367-2630/18/4/045004"
}

@article{Johnson2002,
  author  = "Johnson, Steven G. and Ibanescu, M. and Skorobogatiy, M. A. and Weisberg, O. and Joannopoulos, J. D. and Fink, Y.",
  title   = "Perturbation theory for {Maxwell's} equations with shifting material boundaries",
  journal = "Phys. Rev. E",
  volume  = "65",
  number  = "6",
  pages   = "066611",
  year    = "2002",
  doi     = "10.1103/PhysRevE.65.066611"
}

@article{Abajo2010,
  author  = "{Garc{\'i}a de Abajo}, F. J. and Asenjo-Garcia, A. and Kociak, M.",
  title   = "Multiphoton absorption and emission by interaction of swift electrons with evanescent light fields",
  journal = "Nano {L}ett.",
  volume  = "10",
  number  = "5",
  pages   = "1859--1863",
  year    = "2010",
  doi     = "10.1021/nl100613s"
}

@article{Huang2024,
  author  = "Huang, G. and Beccari, A. and Engelsen, N. J. and Kippenberg, T. J.",
  title   = "Room-temperature quantum optomechanics using an ultralow noise cavity",
  journal = "Nature",
  volume  = "626",
  number  = "7999",
  pages   = "512--516",
  year    = "2024",
  doi     = "10.1038/s41586-023-06997-3"
}

@article{Chen2017,
  author  = "Chen, X. and Chardin, C. and Makles, K. and Ca{\"e}r, C. and Chua, S. and Braive, R. and Robert-Philip, I. and Briant, T. and Cohadon, P.-F. and Heidmann, A. and Jacqmin, T. and Del{\'e}glise, S.",
  title   = "High-finesse {Fabry--Perot} cavities with bidimensional {Si$_3$N$_4$} photonic-crystal slabs",
  journal = "Light {S}ci. {A}ppl.",
  volume  = "6",
  number  = "1",
  pages   = "e16190",
  year    = "2017",
  doi     = "10.1038/lsa.2016.190"
}

@article{Greuter2014,
  author  = "Greuter, L. and Starosielec, S. and Najer, D. and Ludwig, A. and Duempelmann, L. and Rohner, D. and Warburton, R. J.",
  title   = "A small mode volume tunable microcavity: development and characterization",
  journal = "Appl. {P}hys. {L}ett.",
  volume  = "105",
  number  = "12",
  pages   = "121105",
  year    = "2014",
  doi     = "10.1063/1.4896415"
}

@article{Iyer2024,
  author  = "Iyer, A. and Kandel, Y. P. and Xu, W. and Nichol, J. M. and Renninger, W. H.",
  title   = "Coherent optical coupling to surface acoustic wave devices",
  journal = "Nat. {C}ommun.",
  volume  = "15",
  number  = "1",
  pages   = "3993",
  year    = "2024",
  doi     = "10.1038/s41467-024-48167-7"
}

@article{Krause2015,
  author  = "Krause, A. G. and Hill, J. T. and Ludwig, M. and Safavi-Naeini, A. H. and Chan, J. and Marquardt, F. and Painter, O.",
  title   = "Nonlinear radiation pressure dynamics in an optomechanical crystal",
  journal = "Phys. {R}ev. {L}ett.",
  volume  = "115",
  number  = "23",
  pages   = "233601",
  year    = "2015",
  doi     = "10.1103/PhysRevLett.115.233601"
}

@article{Jiang2020,
  author  = "Jiang, W. and Sarabalis, C. J. and Dahmani, Y. D. and Patel, R. N. and Mayor, F. M. and McKenna, T. P. and Van Laer, R. and Safavi-Naeini, A. H.",
  title   = "Efficient bidirectional piezo-optomechanical transduction between microwave and optical frequency",
  journal = "Nat. {C}ommun.",
  volume  = "11",
  number  = "1",
  pages   = "1166",
  year    = "2020",
  doi     = "10.1038/s41467-020-14863-3"
}

@article{Kozak2017,
  author  = "Koz{\'a}k, M. and Beck, P. and Deng, H. and McNeur, J. and Sch{\"o}nenberger, N. and Gaida, C. and Stutzki, F. and Gebhardt, M. and Limpert, J. and Ruehl, A. and Hartl, I. and Solgaard, O. and Harris, J. S. and Byer, R. L. and Hommelhoff, P.",
  title   = "Acceleration of sub-relativistic electrons with an evanescent optical wave at a planar interface",
  journal = "Opt. {E}xpress",
  volume  = "25",
  number  = "16",
  pages   = "19195--19204",
  year    = "2017",
  doi     = "10.1364/OE.25.019195"
}

@article{Henke2021,
  author  = "Henke, J.-W. and Raja, A. S. and Feist, A. and Huang, G. and Arend, G. and Yang, Y. and Kappert, F. J. and Wang, R. N. and M{\"o}ller, M. and Pan, J. and Liu, J. and Kfir, O. and Ropers, C. and Kippenberg, T. J.",
  title   = "Integrated photonics enables continuous-beam electron phase modulation",
  journal = "Nature",
  volume  = "600",
  number  = "7890",
  pages   = "653--658",
  year    = "2021",
  doi     = "10.1038/s41586-021-04197-5"
}

@article{Xi2021,
  author  = "Xi, Xiang and Ma, Jingwen and Zhou, Zhong Hao and Hu, Xin Xin and Chen, Yuan and Zou, Chang Ling and Dong, Chun Hua and Sun, Xiankai",
  title   = "Experimental investigation of the angular symmetry of optical force in a solid dielectric",
  journal = "Optica",
  volume  = "8",
  number  = "11",
  pages   = "1435--1441",
  year    = "2021",
  doi     = "10.1364/OPTICA.423877"
}

@article{Haghighi2018,
  author  = "Moaddel Haghighi, I. and Malossi, N. and Natali, R. and Di Giuseppe, G. and Vitali, D.",
  title   = "Sensitivity--bandwidth limit in a multimode optoelectromechanical transducer",
  journal = "Phys. {R}ev. {A}pplied",
  volume  = "9",
  number  = "3",
  pages   = "034031",
  year    = "2018",
  doi     = "10.1103/PhysRevApplied.9.034031"
}

@article{Vanner2013,
  author  = "Vanner, M. R. and Hofer, J. and Cole, G. D. and Aspelmeyer, M.",
  title   = "Cooling-by-measurement and mechanical state tomography via pulsed optomechanics",
  journal = "Nat. {C}ommun.",
  volume  = "4",
  pages   = "2295",
  year    = "2013",
  doi     = "10.1038/ncomms3295"
}

@article{Esmenda2021,
  author  = "Esmenda, J. C. and Aguila, M. A. C. and Wang, J.-Y. and Lee, T.-H. and Yang, C.-Y. and Lin, K.-H. and Chang-Liao, K.-S. and Katz, N. and Kafanov, S. and Pashkin, Y. A. and Chen, C.-D.",
  title   = "Imaging Off-Resonance Nanomechanical Motion as Modal Superposition",
  journal = "Adv. {S}ci.",
  volume  = "8",
  number  = "13",
  pages   = "2005041",
  year    = "2021",
  doi     = "10.1002/advs.202005041"
}

@article{Tenbrake2024,
  author  = "Tenbrake, L. and Fa{\ss}bender, A. and Hofferberth, S. and Linden, S. and Pfeifer, H.",
  title   = "Direct laser-written optomechanical membranes in fiber {Fabry--Perot} cavities",
  journal = "Nat. {C}ommun.",
  volume  = "15",
  pages   = "209",
  year    = "2024",
  doi     = "10.1038/s41467-023-44490-7"
}

@article{Rochau2021,
  author  = "Rochau, F. and S{\'a}nchez Arribas, I. and Brieussel, A. and Stapfner, S. and Hunger, D. and Weig, E. M.",
  title   = "Dynamical backaction in an ultrahigh-finesse fiber-based microcavity",
  journal = "Phys. {R}ev. {A}pplied",
  volume  = "16",
  number  = "1",
  pages   = "014013",
  year    = "2021",
  doi     = "10.1103/PhysRevApplied.16.014013"
}

@article{Talebi2020,
  author  = "Talebi, N.",
  title   = "Strong Interaction of Slow Electrons with Near-Field Light Visited from First Principles",
  journal = "Phys. {R}ev. {L}ett.",
  volume  = "125",
  number  = "8",
  pages   = "080401",
  year    = "2020",
  doi     = "10.1103/PhysRevLett.125.080401"
}

@article{Ropers2016,
  author  = "Echternkamp, K. E. and Feist, A. and Sch{\"a}fer, S. and Ropers, C.",
  title   = "Ramsey-type phase control of free-electron beams",
  journal = "Nat. {P}hys.",
  volume  = "12",
  pages   = "1000--1004",
  year    = "2016",
  doi     = "10.1038/nphys3844"
}

@article{Joeckel2011,
  author  = "J{\"o}ckel, A. and Rakher, M. T. and Korppi, M. and Camerer, S. and Hunger, D. and Mader, M. and Treutlein, P.",
  title   = "Spectroscopy of mechanical dissipation in micro-mechanical membranes",
  journal = "Appl. {P}hys. {L}ett.",
  volume  = "99",
  number  = "14",
  pages   = "143109",
  year    = "2011",
  doi     = "10.1063/1.3646914"
}

@article{Rossi2017,
  author  = {Rossi, N. and Braakman, F. R. and Cadeddu, D. and Vasyukov, D. and T{\"u}t{\"u}nc{\"u}oglu, G. and
             {Fontcuberta i Morral}, A. and Poggio, M.},
  title   = {Vectorial scanning force microscopy using a nanowire sensor},
  journal = {Nat. {N}anotechnol.},
  volume  = {12},
  pages   = {150--155},
  year    = {2017},
  doi     = {10.1038/nnano.2016.189}
}

@article{Roxworthy2018,
  author  = "Roxworthy, B. J. and Vangara, S. and Aksyuk, V. A.",
  title   = "Subdiffraction spatial mapping of nanomechanical modes using a plasmomechanical system",
  journal = "ACS {P}hotonics",
  volume  = "5",
  number  = "9",
  pages   = "3658--3665",
  year    = "2018",
  doi     = "10.1021/acsphotonics.8b00604"
}

@article{Roxworthy2016,
  author  = {Roxworthy, Brian J. and Aksyuk, Vladimir A.},
  title   = {Nanomechanical motion transduction with a scalable localized gap plasmon architecture},
  journal = {Nat. {C}ommun.},
  volume  = {7},
  pages   = {13746},
  year    = {2016},
  doi     = {10.1038/ncomms13746}
}

@article{Saharian2021,
  author  = {Saharian, A. A. and Mkrtchyan, A. R. and Gevorgian, L. A. and Grigoryan, L. Sh. and Khachatryan, B. V.},
  title   = {Radiation from an electron bunch flying over a surface wave},
  journal = {Nucl. Instrum. Methods Phys. Res. B},
  volume  = {173},
  number  = {1--2},
  pages   = {211--220},
  year    = {2001},
  doi     = {10.1016/S0168-583X(00)00288-3}
}

\bmhead{Acknowledgements}

The authors thank Suraj Kumar for helpful discussions during the early development of this work.

\bmhead{Funding}

This work was supported by the Ministry of Education, Singapore, under the Academic Research Fund Tier 2 program (award no. MOE-T2EP50222-0012).
S.H. was supported by the Singapore International Graduate Award (SINGA).
L.C.K. acknowledges support from the National Research Foundation,
Singapore, and the Ministry of Education, Singapore.

\bmhead{Author contributions}

S.H. conceived the physical idea of the phonon-assisted phase-matching
interaction, developed the theoretical framework, performed the numerical
calculations, analyzed the results, and wrote the original manuscript.
L.W.W. contributed to the physical analysis and discussion, manuscript
writing, and figure preparation. L.J.W. contributed to the initial signal-sensing concept and manuscript
development, reviewed the manuscript, and supervised the work. L.C.K. supervised the overall project, guided the manuscript structure and revisions, and contributed to the development of the signal-sensing theory. All authors reviewed and approved the final
manuscript.

\bmhead{Competing interests}

The authors declare no competing interests.

\end{document}